\documentclass[showpacs,amsmath,amssymb,twocolumn,prl,longbibliography,superscriptaddress]{revtex4-1}

\usepackage{amssymb}
\usepackage[dvips]{graphicx}
\usepackage{enumerate}
\usepackage{epsfig} 
\usepackage{subfigure}
\usepackage{xcolor}
\usepackage[T1]{fontenc}
\usepackage{fullpage}
\usepackage{amsthm,amsfonts,amssymb,amscd,mathrsfs,xspace,framed}
\usepackage{mathrsfs,amsmath}
\usepackage{color}
\usepackage{setspace}
\usepackage{url}
\usepackage{wrapfig}
\usepackage{tikz}
\usepackage{enumitem}
\usepackage{bm}
\usepackage{comment}
\usepackage{txfonts}
\usepackage{array}
\usepackage{color}
\usepackage{braket}
\usepackage{natbib}
\usepackage{cleveref}
\usepackage{graphicx}
\usepackage{pdfpages}

\def\be{\begin{equation}}
\def\ee{\end{equation}}
\def\ba{\begin{eqnarray}}
\def\ea{\end{eqnarray}}

\makeatletter
\AtBeginDocument{\let\LS@rot\@undefined}
\makeatother

\begin{document}

\title{Demonstration of a Reconfigurable Entangled Radiofrequency-Photonic Sensor Network}

\author{Yi Xia$^\#$}

\affiliation{
James C. Wyant College of Optical Sciences, University of Arizona, Tucson, Arizona 85721, USA
}

\author{Wei Li$^\#$}
\email{Current address: Department of Electronic and Information Engineering, Shanxi University, Taiyuan, China}
\affiliation{
Department of Materials Science and Engineering, University of Arizona, Tucson, Arizona 85721, USA
}

\author{William Clark}
\affiliation{
General Dynamics Mission Systems, 8220 East Roosevelt Street, Scottsdale, Arizona 85257, USA}

\author{Darlene Hart}
\affiliation{
General Dynamics Mission Systems, 9 Vreeland Road, Florham Park, New Jersey 07932, USA
}

\author{Quntao Zhuang}
\affiliation{
Department of Electrical and Computer Engineering, University of Arizona, Tucson, Arizona 85721, USA
}
\affiliation{
James C. Wyant College of Optical Sciences, University of Arizona, Tucson, Arizona 85721, USA
}

\author{Zheshen Zhang}
\email{zsz@email.arizona.edu \\ 
$^\#$ Equal contributions}
\affiliation{
Department of Materials Science and Engineering, University of Arizona, Tucson, Arizona 85721, USA
}
\affiliation{
James C. Wyant College of Optical Sciences, University of Arizona, Tucson, Arizona 85721, USA
}

\begin{abstract}
Quantum metrology takes advantage of nonclassical resources such as entanglement to achieve a sensitivity level below the standard quantum limit. To date, almost all quantum-metrology demonstrations are restricted to improving the measurement performance at a single sensor, but a plethora of applications require multiple sensors that work jointly to tackle distributed sensing problems. Here, we propose and experimentally demonstrate a reconfigurable sensor network empowered by continuous-variable (CV) multipartite entanglement. Our experiment establishes a connection between the entanglement structure and the achievable quantum advantage in different distributed sensing problems. The demonstrated entangled sensor network is composed of three sensor nodes each equipped with an electro-optic transducer for the detection of radiofrequency (RF) signals. By properly tailoring the CV multipartite entangled states, the entangled sensor network can be reconfigured to maximize the quantum advantage in distributed RF sensing problems such as measuring the angle of arrival of an RF field. The rich physics of CV multipartite entanglement unveiled by our work would open a new avenue for distributed quantum sensing and would lead to applications in ultrasensitive positioning, navigation, and timing.
\end{abstract}

\maketitle
Quantum metrology~\cite{pirandola2018advances,tan2019resurgence,degen2017quantum,giovannetti2011advances} enables a measurement sensitivity below the standard quantum limit (SQL), as demonstrated in the Laser Interferometer Gravitational-wave Observatory (LIGO)~\cite{abadie2011a,aasi2013enhanced,tse2019,acernese2019}. As a unique quantum resource, entanglement has been utilized to enhance the performance of, e.g., microscopy~\cite{ono2013an}, target detection~\cite{zhang2015entanglement}, and phase estimation~\cite{colangelo2017entanglement}. To date, almost all existing entanglement-enhanced sensing demonstrations operate at a single sensor by entangling the probe with a local reference, but a multitude of applications rely on an array of sensors that work collectively to undertake sensing tasks. On the other hand, entanglement-enhanced optical sensing has been extensively explored, while many useful sensing applications for, e.g., positioning and astronomy operate in the radiofrequency (RF) and microwave spectral ranges. In this regard, quantum illumination enables a signal-to-noise ratio (SNR) advantage over classical schemes in the RF and microwave where ambient noise is abundant~\cite{tan2008quantum,lopaeva2013experimental,zhang2015entanglement,zhuang2017optimum,barzanjeh14microwave,barzanjeh2019experimental}, but quantum illumination has a limited operational range and quantum enhancement~\cite{zhang2015entanglement,barzanjeh14microwave,barzanjeh2019experimental}, due to large diffraction in the microwave and a lack of efficient quantum memories. 

\begin{figure*}
    \centering
    \includegraphics[width=1\textwidth]{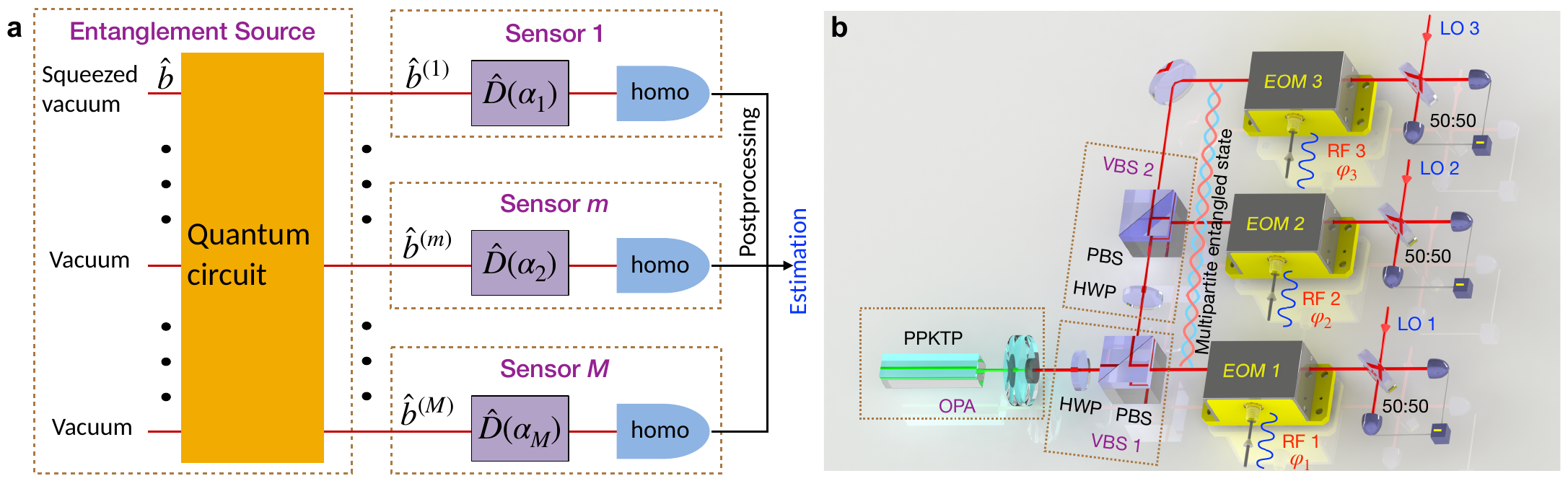}
    \caption{(a) The CV-DQS protocol. A squeezed-vacuum mode $\hat{b}$ is processed by a quantum circuit to produce a multipartite entangled state between $\hat{b}^{(m)}$'s. The sensing process is modeled by displacement operations $\hat{D}(\alpha_m)$'s followed by homodyne measurements (homo), whose outcomes are postprocessed to produce an estimation. (b) Experimental diagram. Phase squeezed light is generated from a periodically-poled KTiOPO$_4$ (PPKTP) crystal embedded in an optical parametric amplifier (OPA) cavity. A quantum circuit comprising two variable BSs (VBSs) configures the CV multipartite entangled state for different distributed sensing tasks. Each VBS consists of a half-wave plate (HWP) and a polarizing BS (PBS). An RF-photonic sensor entails an electro-optic modulator (EOM) and a balanced homodyne detector supplied by a local oscillator (LO) interfering the signal on a 50:50 BS. $\varphi_m$: RF phase.}
    \label{fig:concept_exp}
\end{figure*}

Recent theoretical advances in distributed quantum sensing (DQS) promise a boosted performance for distributed sensing problems. Compared with DQS based on discrete-variable (DV) multipartite entanglement \cite{proctor2018multiparameter,ge2018distributed}, continuous-variable (CV) DQS \cite{zhuang2018distributed,zhuang2019physical} enjoys deterministic preparation of multipartite entangled probe states and robustness against loss. Fig.~\ref{fig:concept_exp}a illustrates CV-DQS: A squeezed vacuum state with a mean photon number $N_S$ is processed by a quantum circuit consisting of beam splitters (BSs) and phase shifters to create a CV multipartite entangled state in $M$ modes, $\{\hat{b}^{(m)}, 1\le m \le M\}$, shared by $M$ sensors. The sensing attempt is modeled by a quadrature displacement operation $\hat{D}(\alpha_m)$ on each mode. The distributed sensing problem is to estimate a global property across all sensors. Let the probed global parameter be $\bar{\alpha}\equiv \sum_{m=1}^M  v_m \alpha_m$, where the weights, $\{ v_m, 1\le m \le M\}$, define the global parameter estimation problem. To estimate the displacements, homodyne measurements yield outcomes $\tilde{\alpha}_m$'s, followed by classical postprocessing that obtains an estimation $\tilde{\alpha}=\sum_m v_m \tilde{\alpha}_m$. 

Critically, the quantum circuit must be optimized to generate a CV multipartite entangled state that minimizes the estimation variance in a given distributed sensing problem. Since $\tilde{\alpha}$ is obtained as if an effective mode $\sum v_m \hat{b}^{(m)}$ is homodyned, the minimum estimation variance is attained when the effective mode equals the original squeezed vacuum mode $\hat{b}$. As such, the optimum quantum circuit distributes $v_m$ amplitude portion of the squeezed vacuum state to the $m$-th sensor, leading to the minimum estimation variance:
\be
\delta\alpha^2 = \frac{\bar{v}^2}{4}\left[\frac{\eta}{\left(\sqrt{N_S+1}+\sqrt{N_S}\right)^2}+1-\eta\right],
\label{dalpha_extension}
\ee
where $\bar{v} \equiv \sqrt{\sum_{m=1}^M v_m^2}$, and $1-\eta$ is the loss at the each sensor. 

Ref.~\cite{xia2019repeater} derived an upper bound for the Fisher information by explicitly reducing it to quadrature variances. The CV-DQS protocol saturates the upper bound in the absence of loss, i.e., $\eta=1$, and is therefore the optimum among {\em all} protocols subject to a photon-number constraint for the probe. Specifically, at $\eta=1$ and a fixed mean photon number $n_s \equiv N_S/M$ at each sensor, equal weights yield $\delta \alpha^2\propto (1/M)^2 \times 1/n_s$, i.e., a Heisenberg scaling for the estimation variance with respect to the number of sensors, whereas any protocol without entanglement is subject to the SQL. Also, $M=1$ and $\bar{v}=1$ reduce the situation to single parameter estimation enhanced by a single-mode squeezed vacuum state. In the presence of loss, the Fisher-information upper bound becomes loose. However, by reducing the multi-parameter estimation problem to a single-parameter estimation problem through a fictitious set of conjugating beamsplitters, it was shown that the CV-DQS protocol remains the optimum among all protocols based on Gaussian states or homodyne measurements~\cite{zhuang2018distributed}. Importantly, the entanglement-enabled advantage in the CV-DQS protocol survives arbitrary amount of loss, even though loss precludes a Heisenberg scaling.

On the experimental front, a recent work~\cite{guo2019distributed} showed that CV entanglement offers a measurement-sensitivity advantage in optical phase estimation over using separable states, but the connection between the entanglement structure and the enabled quantum advantage in different distributed sensing problems have not been experimentally explored. Moreover, the CV-DQS protocol~\cite{zhuang2018distributed} represents a general framework for tackling sensing problems in different physical domains, because quantum transducers can convert the probed physical parameter to quadrature displacement. Here, we demonstrate a reconfigurable entangled sensor network equipped with electro-optic transducers (EOTs) for entanglement-enhanced measurements of RF signals. Our experiment unveils how the CV multipartite entanglement structure determines the quantum advantage in different distributed sensing problems. Specifically, by tailoring the entanglement shared by the sensors, the entangled RF-photonic sensor network achieves an estimation variance 3.2 dB below the SQL in measuring the average RF field amplitudes. Also, in measuring the angle of arrival (AoA) of an emulated incident RF wave, the entangled RF-photonic sensor network achieves an estimation variance 3.2 dB below the SQL via phase-difference estimation at an edge node and 3.5 dB below the SQL via phase-difference estimation at a central node.

The experiment is illustrated in Fig.~\ref{fig:concept_exp}b. Sideband phase squeezed state is generated from an optical parametric amplifier (OPA) and subsequently processed by a quantum circuit comprised of two variable BSs (VBSs) to produce a CV multipartite entangled state shared by three RF-photonic sensors. The RF field at the $m$-th sensor is represented by $\mathcal{E}_m(t) = E_m \cos(\omega_c t + \varphi_m)$, where $\omega_c$ is the carrier frequency, $E_m$ is the amplitude, and $\varphi_m$ is the phase of the RF field. At each sensor, an electro-optic modulator (EOM) driven by the probed RF field induces a displacement on the squeezed phase quadrature, as described by
\begin{eqnarray}
\label{eq:displacement}
    \alpha_m &\simeq& i\sqrt{2}\pi g_m a_c^{(m)} \frac{\gamma E_m}{2V_\pi}  \varphi_m,
\end{eqnarray}
where $g_m=\pm1$ is set by an RF signal delay that controls the sign of the displacement, $a_c^{(m)}$ is the amplitude of the baseband coherent state at the $m$-th sensor, $V_\pi$ is the half-wave voltage of the EOM, and $\gamma$ models the conversion from an external electric field to the internal voltage (see Ref.~\cite{SM}'s Sec.~I~A). To estimate the displacement, a local oscillator (LO) interferes the signal on a 50:50 BS for a balanced homodyne measurement. The time-domain data from the three homodyne measurements are postprocessed to derive the estimated parameter and the associated estimation variance under different settings.

\begin{figure}[t]
    \centering
    \includegraphics[width=0.48\textwidth]{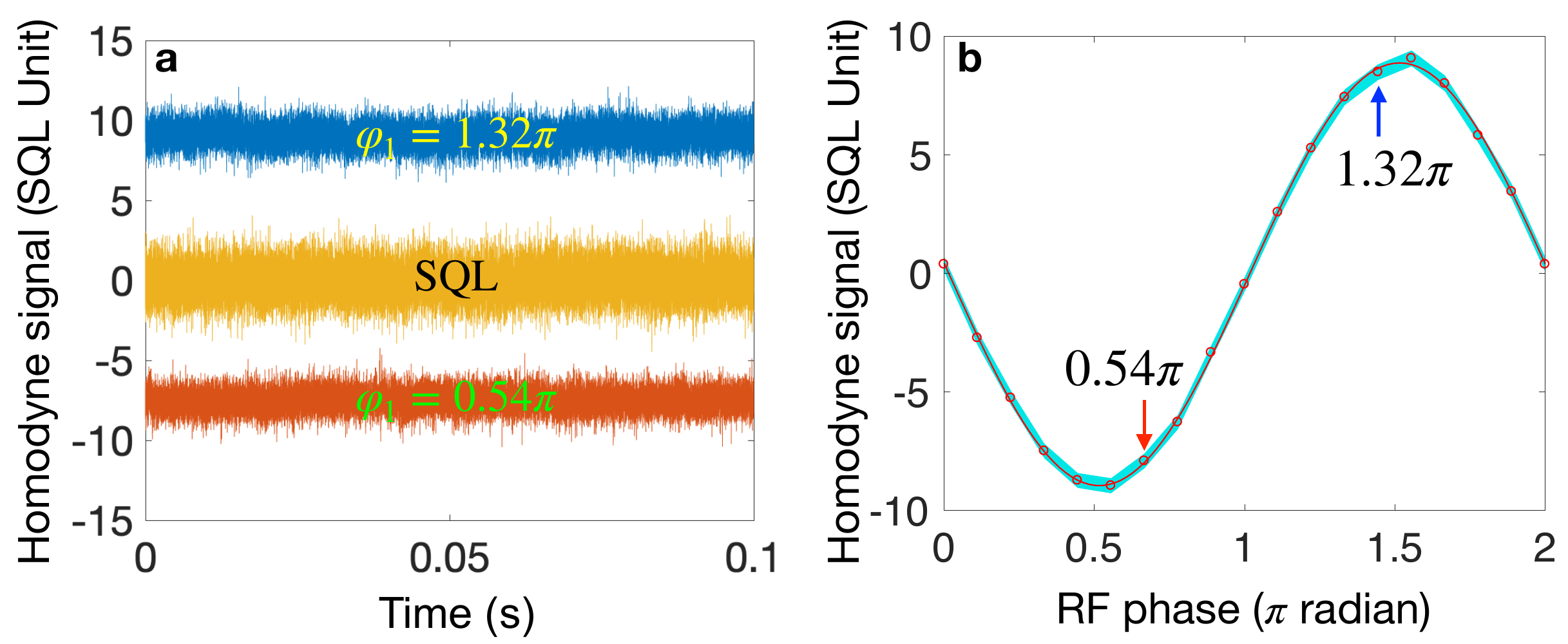}
    \caption{(a) Time-domain traces of the homodyne output signals at $\varphi_1 = 0.54\pi$ (red) and $1.32\pi$ (blue), showing a 4-dB noise reduction as compared with the SQL (gold). The shot-noise limit of the homodyne measurements dictates the SQL and is normalized to a standard deviation of 1. (b) The homodyne output signals at different RF phases. All signals are normalized using the same factor for the SQL normalization. Circles: the means of the measured homodyne signals; red curve: a sinusoidal fit; shaded area: normalized standard deviation of the measurement noise. }
    \label{fig:1-node}
\end{figure}

Prior to constructing an entangled sensor network, we first assess RF-photonic sensing enhanced by single-mode squeezed light. To do so, VBS 1 is configured to deliver all light to Sensor 1. Fig.~\ref{fig:1-node}a first plots in the gold curve a calibrated shot-noise level with standard deviation normalized to 1 to represent the SQL. The blue and red curves are, respectively, the traces at $\varphi_1 = 0.54\pi$ and $1.32\pi$ and are normalized using the same factor as that for the gold curve. The variances of the curves reflect the quantum measurement noise, which in turn determines the estimation variance. Beating the SQL is a nonclassical characteristic, as witnessed in the variances of the red and blue curves. Both cases suppress the SQL by $\sim$ 4 dB. The means of the time-domain homodyne traces, as the phase of the RF field is swept, are then scaled to the SQL unit and plotted as red circles in Fig.~\ref{fig:1-node}b, showing a nice fit to a sinusoidal function, as expected. 

\begin{figure}[t]
    \centering
    \includegraphics[width=.48\textwidth]{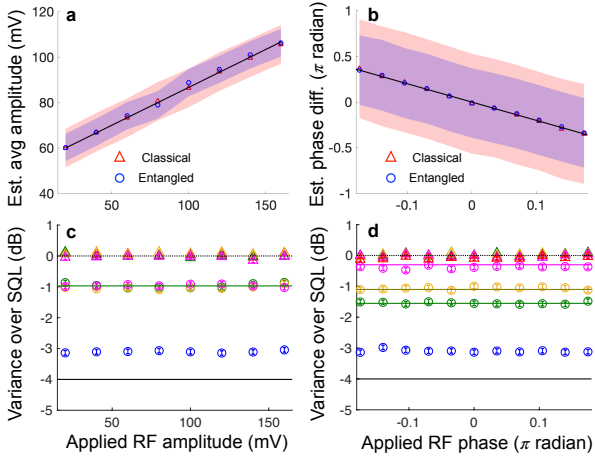}
     \caption{Estimation of (a) average field amplitude and (b) phase difference at an edge node. Circles: data for entangled sensors; triangles: data for classical separable sensors. Shaded area: estimation uncertainties for the entangled (blue) and classical separable (red) sensor networks. The entangled sensors show a clear reduced estimation uncertainty. Noise variances at (c) different field amplitudes and (d) at different phase differences at an edge node. In (c) and (d), measurement noise variances are plotted in green for Sensor 1, gold for Sensor 2, and magenta for Sensor 3. Estimation variances, normalized to the SQL, for entangled (blue) and classical separable (red) sensor networks. Green, gold, and magenta solid horizontal lines: theory curves for noise variances at three sensors. Solid black horizontal line: ideal estimation variance for entangled sensors; experimental deviation likely caused by imperfect phase locking between sensors. Dotted black horizontal line: the SQL for measurement noise variances and normalized estimation variances. While all classical data stay at the SQL, the estimation variances for the entangled sensor networks are sub-SQL and are significantly lower than the measurement noise variances at single sensors. Error bars reflect estimated measurement uncertainties caused by system instabilities.}
     \label{fig:RF_task_data}
\end{figure}

We now demonstrate the power of CV multipartite entanglement in three distributed RF sensing tasks. First, the average RF-field amplitude at the three sensors is estimated using an equally weighted CV multipartite entangled state, which yields the optimum performance (see Ref.~\cite{SM}'s Sec.~I). The RF-field amplitude at Sensor 1 is swept from 20 mV to 160 mV while keep the amplitudes of Sensor 2 and 3 at 80 mV. The homodyne data from the three sensors are first averaged and then scaled to ensure an unbiased estimator. The estimates are plotted as blue circles in Fig.~\ref{fig:RF_task_data}a, with the blue shaded area representing the estimation uncertainty due to quantum measurement noise. The deviation from the linear fit is caused by instability in phase locking. As a comparison, the estimated average RF-field amplitudes by a classical separable sensor network are plotted as red triangles in the same figure, with the red shaded area representing the estimation uncertainty. The entangled sensor network shows a reduced estimation variance of 3.2 dB. 

We then estimate the AoA of an emulated incident RF field. In a one-dimensional sensor array, this sensing problem is translated into the estimation of the phase difference across the sensors, which can be solved by a finite difference method~\cite{SM}. To estimate the phase difference at an edge node (served by Sensor 2 in the experiment), the optimum weights for the CV multipartite entangled state are $[-3/2,2,-1/2]$, generated by setting the splitting ratios (reflectivity:transmissivity) of the VBSs to 50:50 and 75:25. The negative signs in the weights are introduced by adding $\pi$-phase delays at Sensor 2 and Sensor 3. In this measurement, the RF phase at Sensor 1 and 3 is swept from -0.17 rad to 0.17 rad while the RF phase at Sensor 2 is set to 0. The estimated phase difference vs. the applied RF-field phase are plotted in Fig.~\ref{fig:RF_task_data}b for the entangled scheme (blue circles) and compared to that of the classical separable scheme (red triangles) networks with the shaded area representing the estimation uncertainties, showing a 3.2-dB reduction in the estimation variance for the entangled case.

While quantum noise arises in each homodyne measurement, a proper multipartite entangled state leads to a reduction in the overall estimation variance, whereas such an noise reduction mechanism is absent in a classical separable sensor network. Figure~\ref{fig:RF_task_data}c and d show such a behavior. The measurement noise variances at the three sensors, represented by green, gold, and magenta, are plotted for both the entangled (circles) and classical separable (triangles) cases in the SQL unit. To facilitate the comparison, the estimation variances for the three tasks are also normalized to the SQL using a factor predetermined by the weights and are depicted as blue circles and red triangles for, respectively, the entangled and classical separable cases. It is evident that all classical data stay at the SQL. For the entangled sensor network, while the noise variances at each sensor barely suppresses the SQL by $\sim$ 1 dB, the normalized estimation variances, however, substantially beats the SQL by 3.2 $\pm$ 0.1 dB.

\begin{figure}[t]
    \centering
    \includegraphics[width=.4\textwidth]{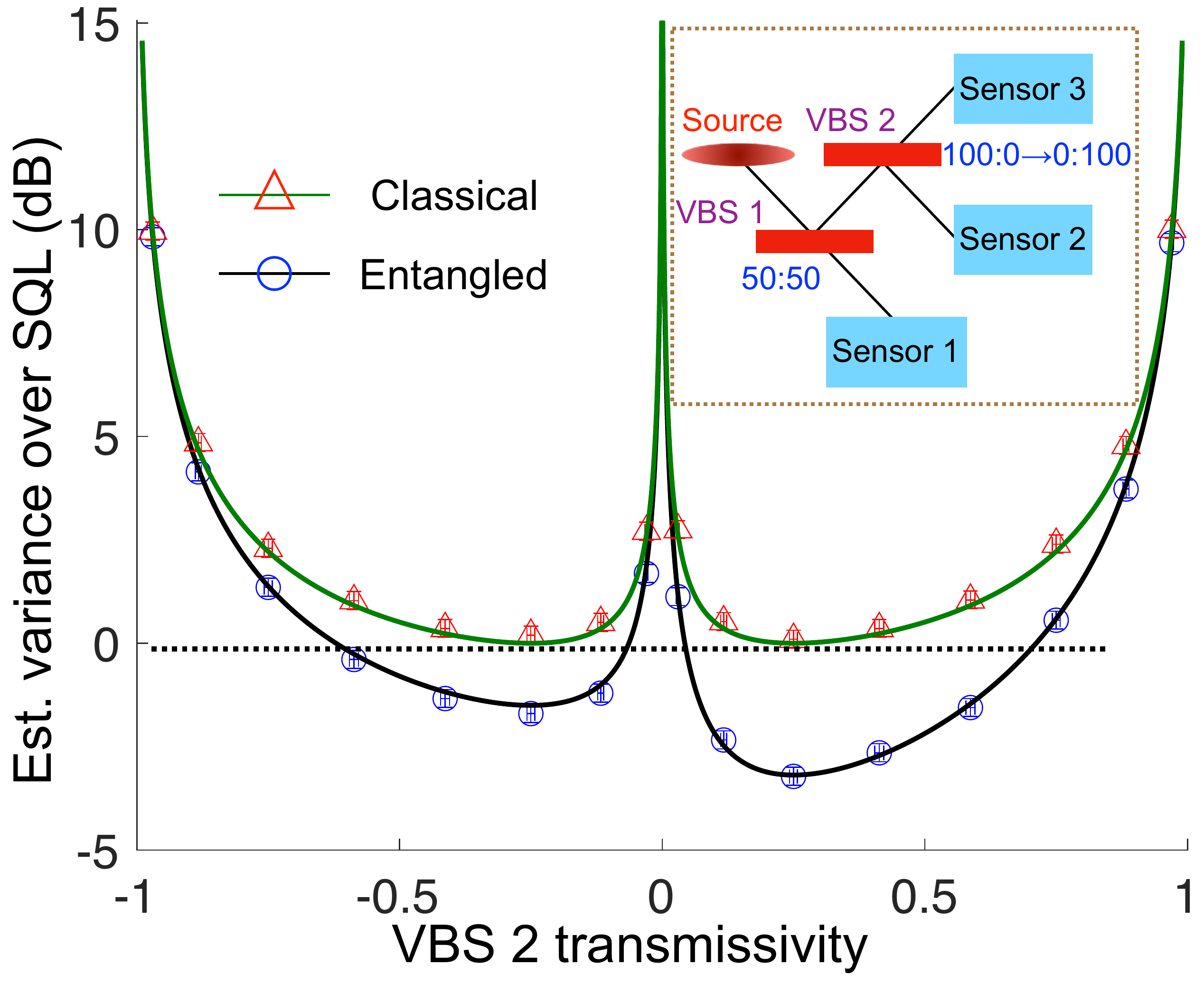}
     \caption{Optimization of CV multipartite entangled state for phase difference at an edge node. Circles: data for entangled sensors; triangles: data for classical separable sensors; black curves: quantum theory; green curves: classical theory. Black horizontal dotted line: the SQL. Error bars account for estimated uncertainties arising from experimental instabilities. The symmetric classical curves vs. the asymmetric quantum curves manifest the correlated quantum noise arising from the homodyne detectors at different sensors. Inset: illustration of the tuning ranges of the splitting ratios for VBSs.}
     \label{fig:optim}
\end{figure}

A unique aspect of an entangled sensor network is that a proper multipartite entangled state need be prepared to achieve the optimum performance in a specific distributed sensing task. To show this, the splitting ratio for VBS 2 is varied to prepare different entangled states for the task of RF-field phase-difference estimation at an edge node (see inset of Fig.~\ref{fig:optim}). The resulting estimation variances are compared with these for the classical separable sensor network under the same VBS settings. In the measurements, a positive transmissivity means the sign of homodyne data remains unchanged in postprocessing while a $\pi$-phase delay is introduced to the RF signal. This is equivalent to applying a $\pi$-phase shift on the quantum state at Sensor 3, followed by a sign flip on its homodyne data. A negative transmissivity indicates a sign flip is applied to Sensor 3's homodyne data in postprocessing and no $\pi$-phase delay on the RF signal. An unbiased estimator is ensured in either case.

The estimation variance vs. transmissivity curves show very different behaviors for the entangled and classical separable cases. The curves for the classical separable case are symmetric, with the minimum estimation variances found at both positive and negative transmissivities, whereas the curves for the entangled case display a strong asymmetric characteristic. Such a behavior manifests the quantum correlation shared by the sensors. In a classical separable sensor network, the quantum measurement noise is independent at different sensors, so postprocessing on the measurement data to acquire an unbiased estimator does not alter the noise power. In an entangled sensor network, however, the quantum measurement noise at different sensors is correlated, so it can only be reduced to minimum if the homodyne data from different sensors are added up with a proper set of weights. Importantly, these weights also need to ensure an unbiased estimator. As such, tailoring a proper CV multipartite entangled state for a specific distributed sensing problem to simultaneously satisfy the two criteria is critical to achieve a large quantum advantage over a classical separable sensor network.

Before closing, a few remarks are worth making. First, our experiment opens a window for quantum-enhanced RF-photonic sensing~\cite{marpaung2019integrated}, which outperforms electronics-based sensing in its large processing bandwidths, engineered RF responses using optical filters, and capability of transporting RF signals over long distances via optical fibers~\cite{lim2010fiber}. A recent photonics-based coherent radar system demonstrated key performance metrics such as a signal-to-noise ratio of 73 dB MHz$^{-1}$ and a spurious-free dynamic range of $70$ dBc, comparable with state-of-the-art electronics-based radar systems' 80 dB MHz$^{-1}$ and $70$ dBc~\cite{ghelfi2014a}. Higher RF-to-photonic conversion efficiency, determined by the $V_\pi$ of the EOT, can further increase the measurement sensitivity. State-of-the-art EOTs based on, e.g., piezo-optomechanical coupling~\cite{jiang2019efficient}, ultrasmall cavities~\cite{choi2017self-similar}, organic EO-plasmonic nanostructures~\cite{heni2019plasmonic}, and highly nonlinear ferroelectric materials~\cite{abel2019large}, can achieve $V_\pi <$ 0.1 V~\cite{jiang2019efficient} and thus increase the measurement sensitivity by > 60 dB. It is worth noting that the quantum advantage survives low RF-photonic conversion efficiency, assuming the same EOTs are employed in both the entangled and classical separable sensor networks. Second, in our experiment, the anti-squeezing level at the source is $\sim$ 10 dB and the squeezing level is $\sim$ 4 dB, from which we can infer the ideal source squeezing of $\sim$ 11.7 dB ($N_S\sim3.3$). The measured squeezing was $\sim$ 3.2 dB for the senor network. Thus, an overall efficiency $\eta\sim 0.56$ is derived. With equal weights, the optimum separable scheme employs $\sim$ 7.9 dB of local squeezing at each sensor~\cite{zhuang2018distributed} to match the total mean photon number and achieves a $2.7$ dB of noise reduction. This leads to a $\sim$ 10\% advantage in estimation variance for our experimental result over that of the optimum separable sensor network, thereby verifying the entanglement shared by the sensors~\cite{qin2019characterizing}. Third, while the current entangled RF-photonic sensor network cannot beat the ultimate estimation precision limit set by the RF sky temperature, it does offer an advantage over a classical RF-photonic sensor network under the same task, assuming sensors are connected by low-loss optical fibers that distribute entanglement over a few kilometers without significant loss penalty~\cite{SM}. To further enlarge the operational range, noiseless linear amplifiers~\cite{ralph2009nondeterministic,xia2019repeater} or CV error correction~\cite{noh2019encoding, zhuang2019distributed} can be used to overcome loss. Fourth, the entangled sensor network does not require quantum memories, but with the assistance of quantum memories it will be able to extract time-domain information more effectively.

In conclusion, we implemented a reconfigurable entangled RF-photonic sensor network and demonstrated sub-SQL estimation variances in distributed sensing tasks. A connection between the entanglement structure and nonclassical quantum noise reduction was established. This quantum-sensing paradigm would create opportunities for ultrasensitive position, navigation, timing, astronomy, and imaging applications.

\begin{acknowledgements}
We thank Jeffrey Shapiro and Franco Wong for their valuable feedback on the manuscript and Saikat Guha for helpful discussions. We also thank an anonymous Referee's comments, which have helped us significantly improve the manuscript. This work is funded by General Dynamics Mission Systems and the Office of Naval Research Award No.~N00014-19-1-2190. We also thank the University of Arizona for the support in the quantum research.
\end{acknowledgements}

{}

\foreach \x in {1,...,11}
{%
\clearpage
\includepdf[pages={\x,{}}]{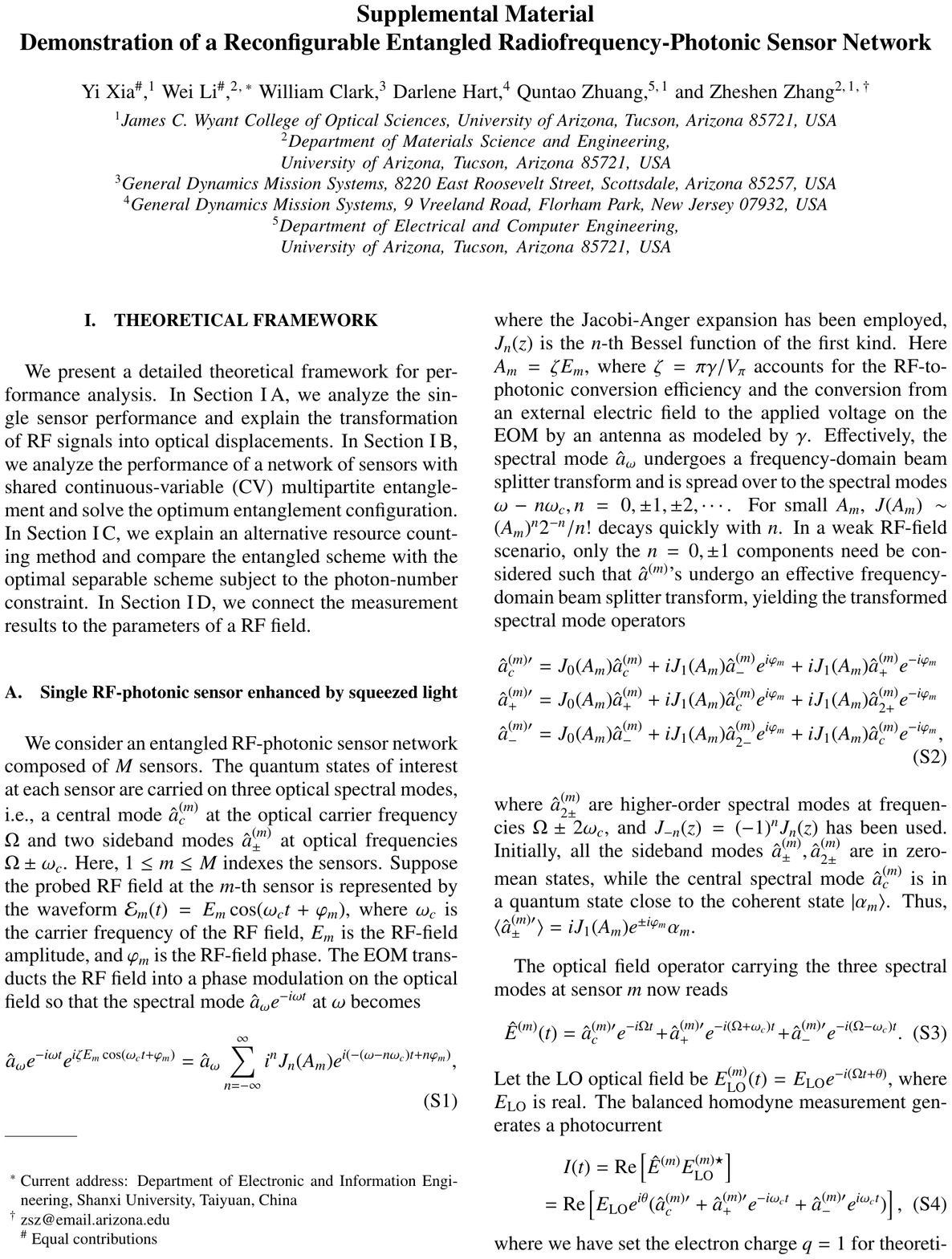}
}



\begin{thebibliography}{10}

\bibitem{pirandola2018advances}
S. Pirandola, B. R. Bardhan, T. Gehring, C. Weedbrook, and S. Lloyd, Advances in photonic quantum sensing, Nat. Photon. {\bf 12}, 724--733 (2018).

\bibitem{degen2017quantum} 
C. L. Degen, F. Reinhard, and P. Cappellaro, Quantum sensing, Rev. Mod. Phys. {\bf 89}, 035002 (2017).

\bibitem{giovannetti2011advances}
V. Giovannetti, S. Lloyd, and L. Maccone, Advances in quantum metrology, Nat. Photon. {\bf 5}, 222 (2011).

\bibitem{tan2019resurgence}
S. H. Tan, and P. P. Rohde, The resurgence of the linear optics quantum interferometer--recent advances \& applications, Rev. Phys. {\bf 4}, 100030 (2019).

\bibitem{abadie2011a} 
J. Abadie {\em et al.} (LIGO Collaboration), A gravitational wave observatory operating beyond the quantum shot-noise limit, Nat. Phys. {\bf 7}, 962--965 (2011).

\bibitem{aasi2013enhanced} 
J. Aasi {\em et al.} (LIGO Collaboration), Enhanced sensitivity of the LIGO gravitational wave detector by using squeezed states of light, Nat. Photon. {\bf 7}, 613--619 (2013).

\bibitem{tse2019}
M. Tse {\em et al.} (LIGO Collaboration), Quantum-Enhanced Advanced LIGO Detectors in the Era of Gravitational-Wave Astronomy, Phys. Rev. Lett. {\bf 123}, 231107 (2019).

\bibitem{acernese2019}
F. Acernese {\em et al}. (Virgo Collaboration), Increasing the Astrophysical Reach of the Advanced Virgo Detector via the Application of Squeezed Vacuum States of Light, Phys. Rev. Lett. {\bf 123}, 231108 (2019).

\bibitem{ono2013an}
T. Ono, R. Okamoto, and S. Takeuchi, An entanglement-enhanced microscope, Nat. Commun. {\bf 4}, 2426 (2013).

\bibitem{zhang2015entanglement}
Z. Zhang, S. Mouradian, F. N. C. Wong, and J. H. Shapiro, Entanglement-enhanced sensing in a lossy and noisy environment, Phys. Rev. Lett. {\bf 114}, 110506 (2015).

\bibitem{colangelo2017entanglement}
G. Colangelo, F. Martin Ciurana, G. Puentes, M. W. Mitchell, and R. J. Sewell, Entanglement-enhanced phase estimation without prior phase information, Phys. Rev. Lett. {\bf 118}, 233603 (2017).

\bibitem{tan2008quantum} 
S.-H. Tan, B. I. Erkmen, V. Giovannetti, S. Guha, S. Lloyd, L. Maccone, S. Pirandola, and J. H. Shapiro, Quantum illumination with Gaussian states, Phys. Rev. Lett. \textbf{101}, 253601 (2008).

\bibitem{lopaeva2013experimental}
E. D. Lopaeva, I. Ruo Berchera, I. P. Degiovanni, S Olivares, G. Brida, and M. Genovese, Experimental realization of quantum illumination, Phys. Rev. Lett. {\bf 110}, 153603 (2013).

\bibitem{zhuang2017optimum} 
Q. Zhuang, Z. Zhang and J. H. Shapiro, Optimum mixed-state discrimination for noisy entanglement-enhanced sensing, Phys. Rev. Lett. {\bf 118}, 040801 (2017).

\bibitem{barzanjeh14microwave}
S. Barzanjeh, S. Guha, C. Weedbrook, D. Vitali, J. H. Shapiro, and S. Pirandola, Microwave quantum illumination, Phys. Rev. Lett. {\bf 114}, 080503 (2015).

\bibitem{barzanjeh2019experimental}
S. Barzanjeh, S. Pirandola, D. Vitali, and J. M. Fink, Experimental microwave quantum illumination, arXiv:1908.03058.

\bibitem{proctor2018multiparameter} 
T. J. Proctor, P. A. Knott, and J. A. Dunningham, Multiparameter estimation in networked quantum sensors, Phys. Rev. Lett. {\bf 120}, 080501 (2018).

\bibitem{ge2018distributed}
W. Ge, K. Jacobs, Z. Eldredge, A. V. Gorshkov, and M. Foss-Feig, Distributed quantum metrology with linear networks and separable inputs, Phys. Rev. Lett. {\bf 121}, 043604 (2018).

\bibitem{zhuang2018distributed} 
Q. Zhuang, Z. Zhang, and J. H. Shapiro, Distributed quantum sensing using continuous-variable multipartite entanglement, Phys. Rev. A {\bf 97}, 032329 (2018).

\bibitem{zhuang2019physical}
Q. Zhuang and Z. Zhang, Physical-layer supervised learning assisted by an entangled sensor network, Phys. Rev. X {\bf 9}, 041023 (2019).

\bibitem{guo2019distributed} 
X. Guo {\em et al.,} Distributed quantum sensing in a continuous variable entangled network, Nat. Phys. {\bf 16}, 281--284(2020).

\bibitem{SM}
See Supplemental Material for details of the theoretical framework, the experimental setup, and additional experimental data. Supplemental Material includes Refs.~\cite{mehmet2011squeezed, qian2019heisenberg, tamura2018the}.

\bibitem{marpaung2019integrated}
D. Marpaung, J. Yao, and J. Capmany, Integrated microwave photonics, Nat. Photon. {\bf 13}, 80--90 (2019).

\bibitem{lim2010fiber}
C. Lim {\em et al.,} Fiber-wireless networks and subsystem technologies, J. Lightw. Technol, {\bf 28}, 390--405 (2010).

\bibitem{ghelfi2014a}
P. Ghelfi {\em et al.,} A fully photonics-based coherent radar system, Nature {\bf 507}, 341--345 (2014).

\bibitem{jiang2019efficient}
W. Jiang {\em et al.}, Efficient bidirectional piezo-optomechanical transduction between microwave and optical frequency, arXiv:1909.04627.

\bibitem{choi2017self-similar}
H. Choi, M. Heuck, and D. Englund, Self-Similar Nanocavity Design with Ultrasmall Mode Volume for Single-Photon Nonlinearities, Phys. Rev. Lett. {\bf 118}, 223605 (2017).

\bibitem{heni2019plasmonic}
W. Heni {\em et al.}, Plasmonic IQ modulators with attojoule per bit electrical energy consumption, Nat. Commun. {\bf 10}, 1694 (2019).

\bibitem{abel2019large}
S. Abel {\em et al.}, Large Pockels effect in micro- and nanostructured barium titanate integrated on silicon, Nat. Mat. {\bf 18}, 42--47(2019).

\bibitem{qin2019characterizing} 
Z. Qin, M. Gessner, Z. Ren, X. Deng, D. Han, W. Li,  X. Su, A. Smerzi, and K. Peng, Characterizing the multipartite continuous-variable entanglement structure from squeezing coefficients and the Fisher information, npj Quantum Inf. {\bf 5}, 3 (2019).

\bibitem{xia2019repeater} 
Y. Xia, Q. Zhuang, W. Clark, and Z. Zhang, Repeater-enhanced distributed quantum sensing based on continuous-variable multipartite entanglement, Phys. Rev. A {\bf 99}, 012328 (2019).

\bibitem{ralph2009nondeterministic} 
T. C. Ralph and A. P. Lund, Nondeterministic noiseless linear amplification of quantum systems, in Proceedings of the Ninth International Conference on Quantum Communication, Measurement and Computing, Calgary, 2008, edited by A. Lvovsky, AIP Conf. Proc. No. 1110 (AIP, Melville, 2009), p. 155.

\bibitem{noh2019encoding}
K. Noh, S. M. Girvin, and L. Jiang, Encoding an oscillator into many oscillators, arXiv:1903.12615.

\bibitem{zhuang2019distributed}
Q. Zhuang, J. Preskill, and L. Jiang, Distributed quantum sensing enhanced by continuous-variable error correction, New J. Phys. {\bf 22}, 022001 (2020).

\bibitem{mehmet2011squeezed}
M. Mehmet, S. Ast, T. Eberle, S. Steinlechner, H. Vahlbruch, and R. Schnabel, Squeezed light at 1550 nm with a quantum noise reduction of 12.3 dB, Opt. Express {\bf 19}, 25763 (2011).

\bibitem{qian2019heisenberg}
K. Qian, Z. Eldredge, W. Ge, G. Pagano, C. Monroe, J. V. Porto, and A. V. Gorshkov, Heisenberg-scaling measurement protocol for analytic functions with quantum sensor networks, Phys. Rev. A {\bf 100}, 042304 (2019).

\bibitem{tamura2018the}
Y. Tamura, H. Sakuma, K. Morita, M. Suzuki, Y. Yamamoto, K. Shimada, Y. Honma, K. Sohma, T. Fujii, and T. Hasegawa, The First 0.14-dB/km Loss Optical Fiber and its Impact on Submarine Transmission, J. Light. Technol. {\bf 36}, 44--49 (2018).


\end{thebibliography}
\end{document}